# A brown dwarf mass donor in an accreting binary

S.P. Littlefair[1], V.S. Dhillon[1], T.R. Marsh[2], Boris T. Gänsicke[2], John Southworth[2], C.A. Watson[1]

[1]*Dept of Physics and Astronomy, University of Sheffield, S3 7RH, UK*

[2]*Dept of Physics, University of Warwick, Coventry, CV4 7AL, UK*

**A long standing and unverified prediction of binary star evolution theory is the existence of a population of white dwarfs accreting from sub-stellar donor stars. Such systems ought to be common, but the difficulty of finding them, combined with the challenge of detecting the donor against the light from accretion means that no donor star to date has a measured mass below the hydrogen burning limit. Here we apply a technique which allows us to reliably measure the mass of the unseen donor star in eclipsing systems. We are able to identify a brown dwarf donor star, with a mass of 0.052±0.002 M$_\odot$. The relatively high mass of the donor star for its orbital period suggests that current evolutionary models may underestimate the radii of brown dwarfs.**

The theory of binary star evolution invokes core astrophysics, including stellar models, magnetic braking and gravitational radiation. Because a large fraction of all stars are found in binaries *(1),* and because the predictions of binary evolution theory describe some of the most exotic objects in our Universe, including the likely progenitors of short γ-ray bursts *(2)* and type Ia supernovae *(3),* and how they may evolve with time, the study of binary star evolution has wide ranging impact throughout astronomy and cosmology. It is therefore a cause of serious concern that the predictions of binary star evolution theory are, in some cases, dramatically out of line with observations. A prime example is the apparent lack of brown dwarf donor stars amongst the binaries known as cataclysmic variables (CVs). CVs are short-period (typically, P$_{orb}$

< 1d) binaries containing a white dwarf primary star and a low-mass donor star. The donor star is so close to the white dwarf that it is tidally distorted and fills a critical surface known as the Roche Lobe, which determines the maximum extent of a star in a close binary. The secular evolution of CVs is driven by angular momentum loss due to gravitational radiation, magnetic braking of the donor star and perhaps circumbinary discs *(4)*. The removal of angular momentum from the binary drives mass transfer from the donor star to the white dwarf, via an accretion disc. The donor shrinks as it loses mass, causing the orbital period to decrease. This continues until the donor's mass drops below the hydrogen burning limit, at which point the donor star becomes a brown dwarf. The resulting changes in the donor's internal structure means that it now expands in response to mass loss, causing the orbital period to increase *(5)*. Thus, CVs are expected to show a minimum orbital period, and those CVs which have evolved past the period minimum (post-period minimum systems) should possess brown dwarf donor stars. Theoretical studies *(6, 7)* predict that around 70% of the current CV population have evolved past the orbital period minimum. However, despite extensive observational effort *(8-15)*, not one of the ~1600 known CVs has a donor which has been unambiguously shown to be sub-stellar *(8)*.

Whilst there has been speculation that the rate of angular momentum loss is so low that systems may not have had time to reach their minimum period *(16)*, or that it is enhanced by circumbinary discs to rates so high that the donor is rapidly devoured *(17)*, it may be that the observed lack of post-period minimum systems is a result of selection effects. Post-period minimum systems will have low mass transfer rates and will consequently be very faint. They may also lack the frequent outbursts which aid in identifying their younger counterparts *(18)*. Even if post-period minimum systems do form part of the known CV population, direct detection of the donor star is extremely difficult against the background of the relatively bright white dwarf and accretion disc *(8)*.

Recent developments have allowed these problems to be overcome. The Sloan Digital Sky Survey (SDSS) *(19-23)* goes much fainter than previous surveys, and as objects are selected on the basis of their spectra, CVs need not show outbursts to be included. The SDSS sample could therefore contain a large number of post-period minimum systems. Whilst direct detection of the donor star in these systems remains a challenge, it is possible to measure the mass and radius of the donor in eclipsing CVs. By fitting a simple physical model (see supporting online material for details) to the eclipse light curve it is possible to obtain a full solution of the geometrical and physical parameters of the binary, and in particular the masses of the white dwarf and donor *(24, 25)*. Only three assumptions are made: that the matter transferred between the donor and white dwarf follows a ballistic trajectory until it impacts the outer edge of the accretion disc; that the white dwarf follows a theoretical mass-radius relation; and that the donor fills its Roche Lobe.

We applied this method to the short-period CV SDSS 103533.03+055158.4 (hereafter SDSS 1035). After discovery within the SDSS *(23)*, our own follow-up Very Large Telescope spectroscopy *(26)* found the system to be eclipsing. We obtained high time-resolution photometry of 8 eclipses between 4-8$^{th}$ March 2006, using Ultracam on the 4.2-m William Herschel Telescope. Ultracam provides simultaneous photometry in the Sloan-*u'g'r'* colour filters with minimal dead-time between exposures. Mid-eclipse times were calculated by averaging the white dwarf ingress and egress times, which are given by the minimum and maximum of the light curve derivative respectively *(25)*. The orbital ephemeris was found with a linear least-squares fit to the times of mid-eclipse, giving an orbital period of 82.0896 ± 0.0003 minutes. The 8 eclipse light-curves were phased according to our ephemeris, averaged and then binned by five data points to produce an average light curve for each band (Fig. 1). Sharp steps in the light curves represent the ingress and egress of the white dwarf behind the donor. The white dwarf eclipse is symmetric around binary phase 0, with ingress and egress near phase -0.02 and 0.02 respectively. Also visible is the eclipse of the bright spot, where the gas stream hits the outer edge of the accretion disc. Bright spot ingress is visible near phase 0.01, with egress near phase 0.08. The presence of a bright spot confirms ongoing accretion, validating our assumption that the donor fills its Roche Lobe. The average light curves in each band were fitted separately with a geometric model including a limb-darkened white dwarf and a bright spot modelled as a linear strip passing through the intersection of the gas stream and accretion disc (full details are contained in the supporting online material). The model results are combined with a theoretical white dwarf mass-radius relationship to obtain a full solution for the binary parameters (Table 1).

The most important result is the donor's mass, $M_c = 0.052 \pm 0.002\ M_\odot$. This is comfortably below the hydrogen burning limit of around $\sim 0.072\ M_\odot$ for solar metallicities *(27)*, making the donor star in SDSS 1035 a confirmed brown-dwarf in a CV; only one other is known in any accreting binary system *(28)*. This discovery supports a fundamental and long-standing prediction of binary evolution theory; that a population of post-period minimum CVs exists, thus refuting claims that binary evolution may be too slow for such systems to form *(16)*. It also demonstrates that the SDSS CV survey is sensitive to post-period minimum systems; the spectroscopic properties of SDSS 1035 are not unusual for the short period CVs found within the SDSS *(23)* and therefore, if the population synthesis models *(6, 7)* are correct, the SDSS CV sample should contain large numbers of post-period minimum systems.

It is possible, though unlikely, that SDSS 1035 could have formed directly from a detached white dwarf/brown dwarf binary similar to WD0137-349 *(29)*. The progenitors of such systems are solar-type stars with brown dwarf companions at separations of a few au *(30)*; such binaries fall within the "brown-dwarf desert" and are very rare *(31)* and so only a few percent of CVs should form from binaries like WD0137-349 *(29)*. It is therefore much more likely that SDSS 1035 is indeed a post-period minimum CV.

Even if SDSS 1035 formed from a white dwarf/brown dwarf binary, its existence shows that an accreting white dwarf/brown dwarf binary is a viable configuration. Since the secular evolution of CVs moves them towards this configuration, this makes the existence of post-period minimum CVs highly probable.

The white dwarf temperature, derived from the colours of the white dwarf eclipse (supporting online material), can be used to determine the long-term average of the mass transfer rate *(32)*. We find a mass transfer rate of $(10 \pm 2) \times 10^{-12}$ $M_\odot$ yr$^{-1}$, in line with the predictions from gravitational radiation, but inconsistent with predictions that include a circumbinary disc, in which the mass transfer rate is increased to $80 \times 10^{-12}$ $M_\odot$ yr$^{-1}$. Increased angular momentum loss due to circumbinary discs is invoked to explain many problems in binary evolution, including the discrepancy between the observed and predicted values of the minimum orbital period for CVs and the apparent lack of large numbers of post-period minimum systems *(4, 17)*. The low inferred mass transfer rate in SDSS 1035, however, argues against models including circumbinary discs to explain these discrepancies.

A comparison of the donor mass in SDSS 1035 to current evolutionary models *(4)* is shown in Fig. 2. We can see that the mass of the donor in SDSS 1035 is inconsistent with models where gravitational radiation is the sole source of angular momentum loss. The donor mass is consistent with models including a circumbinary disc, but these models are ruled out by the inferred mass transfer rate. The discrepancy between observed and predicted masses is probably not due to the rapid rotation and/or distortion of the donor *(33)*, but might be due to irradiation from the white dwarf, or nuclear evolution of the progenitor star *(34)*. Alternatively, the source of the discrepancy may lie with current stellar models, which are based on an up-to-date equation of state specifically calculated for very low mass stars, brown dwarfs and giant planets *(35)*. For the donor star in SDSS 1035 to fill its Roche Lobe implies the radius must be larger than predicted by ~10%. If current models do underestimate the radii of brown dwarfs, this implies that the inferred ages and masses for isolated brown dwarfs are in error. Additional theoretical work will be necessary to determine if any or all of these ideas are sufficient to explain the discrepancy between the observed and predicted mass and radius presented here.

36. We thank U. Kolb and I. Baraffe for productive discussions. SPL, CAW and JS are supported by the Particle Physics and Astronomy Research Council (PPARC). TRM acknowledges the support of a PPARC Senior Research Fellowship. BTG acknowledges the support of a PPARC advanced fellowship. Ultracam is supported by PPARC.

Correspondence and requests for materials should be addressed to S.P.L. (s.littlefair@shef.ac.uk).

**Figure 1: Eclipse light curves and model fits for SDSS 1035 a.** The phase folded *u'* light curve. **b.** The phase folded *g'* light curve. **c.** The phase folded *r'* light. Each light curve is fitted separately using the model described in the Supporting Online Material. The data (black) are shown with the fit (red) overlaid, and the residuals plotted below (black). Also shown are the separate light curves of the white dwarf (blue), bright spot (green), accretion disc (purple) and the donor star (orange). Data points excluded from the fit are shown in red.

**Figure 2: Normalized probability distribution functions (PDFs) of the present-day CV population in the ($M_c$, $P_{orb}$) plane (adapted from ref 4).**

**a.** Evolutionary tracks with angular momentum loss driven by gravitational radiation. **b.** Evolutionary tracks with additional angular momentum loss from a circumbinary disc. The present-day CV population is obtained by weighting the contribution of each system to the PDF according to the accretion luminosity, as $L_{acc}^{1.5}$. In each panel an inset is displayed, showing the location of SDSS 1035 in the ($M_c$, $P_{orb}$) plane.

**Table 1 Derived Parameters of SDSS 1035**

| | |
|---|---|
| Mass Ratio $q$ | 0.055±0.002 |
| Inclination $i$ | $83°.1±0°.2$ |
| Orbital Separation $a$ | 0.622±0.003 $R_\odot$ |
| White Dwarf Mass $M_w$ | 0.94±0.01 $M_\odot$ |
| White Dwarf Radius $R_w$ | 0.0087±0.0001 $R_\odot$ |
| White Dwarf Temperature $T^{eff}_w$ | 10100±200 K |
| Donor Star Mass $M_c$ | 0.052±0.002 $M_\odot$ |
| Donor Star Radius $R_c$ | 0.108±0.003 $R_\odot$ |
| Disc Radius $R_d/a$ | 0.362±0.003 |

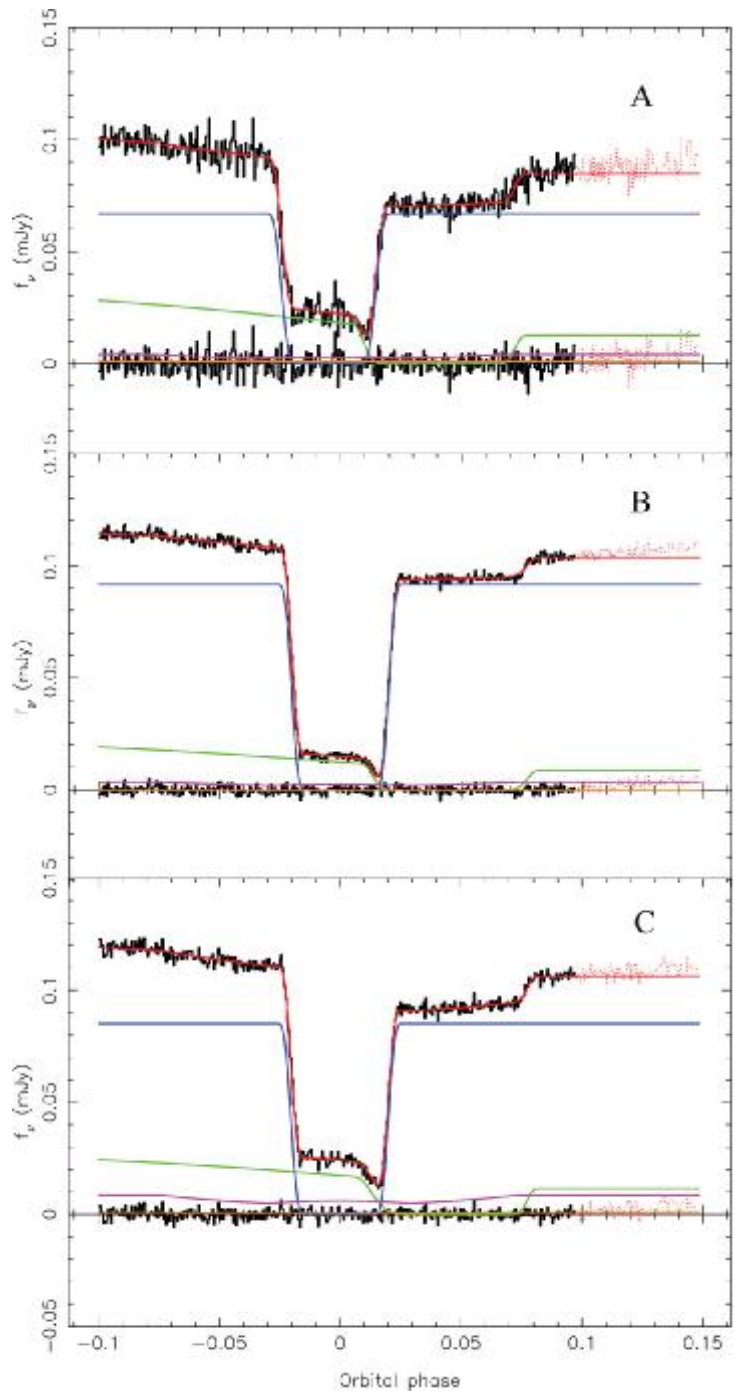

Figure 1

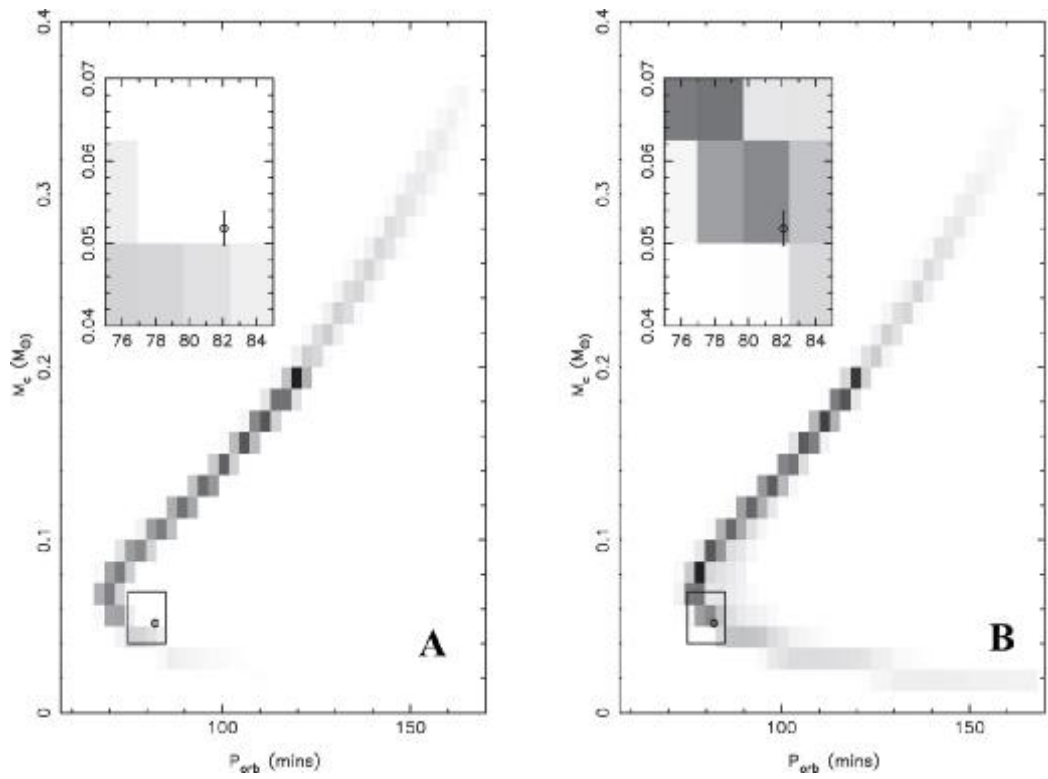

Figure 2.